
\input harvmac

\Title{}
{\vbox{\centerline{On Uplifted SUSY-Breaking Vacua and Direct Mediation}
\bigskip
\centerline{in Generalized SQCD}
}}
\bigskip

\centerline{\it Roberto Auzzi, Shmuel Elitzur and Amit Giveon}
\bigskip
\smallskip
\centerline{Racah Institute of Physics, The Hebrew University}
\centerline{Jerusalem 91904, Israel.}
\smallskip

\vglue .3cm

\bigskip

\let\includefigures=\iftrue
\bigskip
\noindent We search for viable models of direct gauge mediation,
where the SUSY-breaking sector is (generalized) SQCD, which has
cosmologically favorable uplifted vacua even when the reheating
temperature is well above the messenger scale. This requires a
relatively large tadpole term in the scalar potential for the
spurion field $X$ and, consequently, we argue that pure (deformed)
SQCD is not a viable model. On the other hand, in SQCD with an
adjoint, which is natural e.g. in string theory, assuming an
appropriate sign in the K\"ahler potential for $X$, such
metastable vacua are possible.

\bigskip

\Date{January 2010}

\lref\GiudiceBP{
  G.~F.~Giudice and R.~Rattazzi,
  ``Theories with gauge-mediated supersymmetry breaking,''
  Phys.\ Rept.\  {\bf 322}, 419 (1999)
  [arXiv:hep-ph/9801271].
}

\lref\IntriligatorCP{
  K.~A.~Intriligator and N.~Seiberg,
  ``Lectures on Supersymmetry Breaking,''
  Class.\ Quant.\ Grav.\  {\bf 24}, S741 (2007)
  [arXiv:hep-ph/0702069].
}

\lref\IntriligatorDD{
  K.~A.~Intriligator, N.~Seiberg and D.~Shih,
  ``Dynamical SUSY breaking in meta-stable vacua,''
  JHEP {\bf 0604}, 021 (2006)
  [arXiv:hep-th/0602239].
}

\lref\IntriligatorPY{
  K.~A.~Intriligator, N.~Seiberg and D.~Shih,
  ``Supersymmetry Breaking, R-Symmetry Breaking and Metastable Vacua,''
  JHEP {\bf 0707}, 017 (2007)
  [arXiv:hep-th/0703281].
}

\lref\GiveonEF{
  A.~Giveon and D.~Kutasov,
  ``Stable and Metastable Vacua in SQCD,''
  Nucl.\ Phys.\  B {\bf 796}, 25 (2008)
  [arXiv:0710.0894 [hep-th]].
}

\lref\KomargodskiJF{
  Z.~Komargodski and D.~Shih,
  ``Notes on SUSY and R-Symmetry Breaking in Wess-Zumino Models,''
  JHEP {\bf 0904}, 093 (2009)
  [arXiv:0902.0030 [hep-th]].
}

\lref\GiveonYU{
  A.~Giveon, A.~Katz and Z.~Komargodski,
  ``Uplifted Metastable Vacua and Gauge Mediation in SQCD,''
  JHEP {\bf 0907}, 099 (2009)
  [arXiv:0905.3387 [hep-th]].
}

\lref\KitanoXG{
  R.~Kitano, H.~Ooguri and Y.~Ookouchi,
  ``Direct mediation of meta-stable supersymmetry breaking,''
  Phys.\ Rev.\  D {\bf 75}, 045022 (2007)
  [arXiv:hep-ph/0612139].
}

\lref\habamaru{
N.~Haba and N.~Maru,
``A Simple Model of Direct Gauge Mediation of
Metastable Supersymmetry Breaking,''
     Phys.\ Rev.\  D {\bf 76}, 115019 (2007)
     [arXiv:0709.2945 [hep-ph]].
}

\lref\GiveonEW{
  A.~Giveon and D.~Kutasov,
  ``Stable and Metastable Vacua in Brane Constructions of SQCD,''
  JHEP {\bf 0802}, 038 (2008)
  [arXiv:0710.1833 [hep-th]].
}

\lref\ZurZG{
  B.~K.~Zur, L.~Mazzucato and Y.~Oz,
  ``Direct Mediation and a Visible Metastable Supersymmetry Breaking Sector,''
  JHEP {\bf 0810}, 099 (2008)
  [arXiv:0807.4543 [hep-ph]].
}

\lref\GiveonUR{
  A.~Giveon, D.~Kutasov, J.~McOrist and A.~B.~Royston,
  ``D-Terms and Supersymmetry Breaking from Branes,''
  Nucl.\ Phys.\  B {\bf 822}, 106 (2009)
  [arXiv:0904.0459 [hep-th]].
}

\lref\KoschadeQU{
  D.~Koschade, M.~McGarrie and S.~Thomas,
  ``Direct Mediation and Metastable Supersymmetry Breaking for SO(10),''
  arXiv:0909.0233 [hep-ph].
}

\lref\AbelZE{
  S.~A.~Abel, J.~Jaeckel and V.~V.~Khoze,
  ``Gaugino versus Sfermion Masses in Gauge Mediation,''
  arXiv:0907.0658 [hep-ph].
}

\lref\BarnardIR{
  J.~Barnard,
  ``Tree Level Metastability and Gauge Mediation in Baryon Deformed SQCD,''
  arXiv:0910.4047 [hep-ph].
}

\lref\GiveonSR{
  A.~Giveon and D.~Kutasov,
  ``Brane dynamics and gauge theory,''
  Rev.\ Mod.\ Phys.\  {\bf 71}, 983 (1999)
  [arXiv:hep-th/9802067].
}

\lref\GMR{
A.~Giveon, J.~McOrist and A.~B.~Royston,
in progress.
}

\lref\GiveonFK{
  A.~Giveon and D.~Kutasov,
  ``Gauge symmetry and supersymmetry breaking from intersecting branes,''
  Nucl.\ Phys.\  B {\bf 778}, 129 (2007)
  [arXiv:hep-th/0703135].
}

\lref\AmaritiVK{
  A.~Amariti, L.~Girardello and A.~Mariotti,
  ``Non-supersymmetric meta-stable vacua in SU(N) SQCD with adjoint matter,''
  JHEP {\bf 0612}, 058 (2006)
  [arXiv:hep-th/0608063].
}

\lref\AbelCR{
  S.~A.~Abel, C.~S.~Chu, J.~Jaeckel and V.~V.~Khoze,
  ``SUSY breaking by a metastable ground state: Why the early universe
  preferred the non-supersymmetric vacuum,''
  JHEP {\bf 0701}, 089 (2007)
  [arXiv:hep-th/0610334].
}

\lref\CraigKX{
  N.~J.~Craig, P.~J.~Fox and J.~G.~Wacker,
  ``Reheating metastable O'Raifeartaigh models,''
  Phys.\ Rev.\  D {\bf 75}, 085006 (2007)
  [arXiv:hep-th/0611006].
}

\lref\FischlerXH{
  W.~Fischler, V.~Kaplunovsky, C.~Krishnan, L.~Mannelli and M.~A.~C.~Torres,
  ``Meta-Stable Supersymmetry Breaking in a Cooling Universe,''
  JHEP {\bf 0703}, 107 (2007)
  [arXiv:hep-th/0611018].
}

\lref\AbelMY{
  S.~A.~Abel, J.~Jaeckel and V.~V.~Khoze,
  ``Why the early universe preferred the non-supersymmetric vacuum. II,''
  JHEP {\bf 0701}, 015 (2007)
  [arXiv:hep-th/0611130].
}

\lref\KatzGH{
  A.~Katz,
  ``On the Thermal History of Calculable Gauge Mediation,''
  JHEP {\bf 0910}, 054 (2009)
  [arXiv:0907.3930 [hep-th]].
}

\lref\SeibergPQ{
  N.~Seiberg,
  ``Electric - magnetic duality in supersymmetric nonAbelian gauge theories,''
  Nucl.\ Phys.\  B {\bf 435}, 129 (1995)
  [arXiv:hep-th/9411149].
}

\lref\DolanQD{
  L.~Dolan and R.~Jackiw,
  ``Symmetry Behavior At Finite Temperature,''
  Phys.\ Rev.\  D {\bf 9}, 3320 (1974).
}

\lref\KutasovVE{
  D.~Kutasov,
  ``A Comment on duality in N=1 supersymmetric nonAbelian gauge theories,''
  Phys.\ Lett.\  B {\bf 351}, 230 (1995)
  [arXiv:hep-th/9503086].
}

\lref\KutasovNP{
  D.~Kutasov and A.~Schwimmer,
  ``On duality in supersymmetric Yang-Mills theory,''
  Phys.\ Lett.\  B {\bf 354}, 315 (1995)
  [arXiv:hep-th/9505004].
}

\lref\KutasovSS{
  D.~Kutasov, A.~Schwimmer and N.~Seiberg,
  ``Chiral Rings, Singularity Theory and Electric-Magnetic Duality,''
  Nucl.\ Phys.\  B {\bf 459}, 455 (1996)
  [arXiv:hep-th/9510222].
}

\lref\CraigHF{
  N.~Craig, R.~Essig, S.~Franco, S.~Kachru and G.~Torroba,
  ``Dynamical Supersymmetry Breaking, with Flavor,''
  arXiv:0911.2467 [hep-ph].
}

\lref\EssigKZ{
  R.~Essig, J.~F.~Fortin, K.~Sinha, G.~Torroba and M.~J.~Strassler,
  ``Metastable supersymmetry breaking and multitrace deformations of SQCD,''
  JHEP {\bf 0903}, 043 (2009)
  [arXiv:0812.3213 [hep-th]].
}

\lref\GiveonBV{
  A.~Giveon, D.~Kutasov and O.~Lunin,
  ``Spontaneous SUSY Breaking in Various Dimensions,''
  Nucl.\ Phys.\  B {\bf 822}, 127 (2009)
  [arXiv:0904.2175 [hep-th]].
}

\lref\KutasovKB{
  D.~Kutasov, O.~Lunin, J.~McOrist and A.~B.~Royston,
  ``Dynamical Vacuum Selection in String Theory,''
  arXiv:0909.3319 [hep-th].
}

\lref\ArkaniHamedFQ{
  N.~Arkani-Hamed, M.~A.~Luty and J.~Terning,
  ``Composite quarks and leptons from dynamical supersymmetry breaking  without
  messengers,''
  Phys.\ Rev.\  D {\bf 58}, 015004 (1998)
  [arXiv:hep-ph/9712389].
}

\lref\LutyVR{
  M.~A.~Luty and J.~Terning,
  ``Improved single sector supersymmetry breaking,''
  Phys.\ Rev.\  D {\bf 62}, 075006 (2000)
  [arXiv:hep-ph/9812290].
}

\lref\CheungES{
  C.~Cheung, A.~L.~Fitzpatrick and D.~Shih,
  ``(Extra)Ordinary Gauge Mediation,''
  JHEP {\bf 0807}, 054 (2008)
  [arXiv:0710.3585 [hep-ph]].
}

\lref\duncan{
M.~J.~Duncan and L.~G.~Jensen,
  ``Exact tunneling solutions in scalar field theory,''
  Phys.\ Lett.\  B {\bf 291}, 109 (1992).
}

\lref\KatzGV{
  A.~Katz, Y.~Shadmi and T.~Volansky,
  ``Comments on the meta-stable vacuum in N(f) = N(c) SQCD and direct
  mediation,''
  JHEP {\bf 0707}, 020 (2007)
  [arXiv:0705.1074 [hep-th]].
}

\newsec{Introduction}

If supersymmetry (SUSY) provides an explanation to the hierarchy between the weak scale
and the Planck scale, its discovery is just around the corner.
Breaking supersymmetry entirely in the Minimal Supersymmetric extension of
the Standard Model (MSSM) is excluded by experiment.
Thus, in addition to the MSSM we must have a SUSY-breaking sector,
and SUSY breaking is mediated by messengers
to the MSSM (for a review, see e.g. \GiudiceBP).
Especially appealing models are those of direct mediation, where the messengers
are part of the SUSY-breaking sector.

Supersymmetry must be broken spontaneously,
but to explain the hierarchy it should better be broken dynamically,
e.g. in Supersymmetric QCD (SQCD) and its generalizations.
Moreover, to obtain non-vanishing gaugino masses, R-symmetry must be broken
and, consequently, metastable states are highly favored phenomenologically over global minima,
in the landscape of vacua of SQCD (for a review, see e.g. \IntriligatorCP).

Interestingly, metastable vacua are generic in SQCD, as was shown recently in \IntriligatorDD.
We shall refer to metastable states of the type found in \IntriligatorDD\ as the ISS-like vacua.
There is however a serious problem with ISS-like vacua: even when R-symmetry is broken
(e.g. explicitly as in \refs{\IntriligatorPY,\habamaru,\GiveonEF}),
the gaugino masses remain anomalously small
relative to the soft scalar masses, leading to the undesired
necessity of a high fine tuning of the Higgs mass.

The generality of this phenomenon was understood very recently in \KomargodskiJF,
and its important consequence is the following (see also \GiveonYU).
{}For supersymmetry to provide an explanation to the hierarchy problem,
we are likely to live in an {\it uplifted} metastable vacuum
(of the type studied e.g. in \refs{\KitanoXG,\GiveonEW,\ZurZG,\GiveonUR}).
By an uplifted vacuum we mean a SUSY-breaking metastable one,
which is part of an approximate pseudo-moduli space,
along which some messengers are unstable at some other point.
The ISS-like metastable vacua are {\it not} uplifted.

Yet, the existence of uplifted vacua was
shown to be a generic feature of (deformed) SQCD in \GiveonYU,
and such vacua were used for direct gauge mediation of SUSY breaking
in \refs{\GiveonYU,\KoschadeQU}.
However, the parameters space of
the examples in \GiveonYU\ is rather constrained,
and more generic phenomenologically viable
uplifted vacua in the landscape of SQCD
are desired.~\foot{Other types of uplifted
vacua were studied more recently, e.g. in \refs{\AbelZE,\BarnardIR}.}

The purpose of this work is to investigate the landscape of SQCD
in order to identify uplifted metastable vacua which can be used
as the gauge mediation minima in models with interesting
phenomenology. Concretely, we inspect models with a non-trivial
K\"ahler potential for the spurion of SUSY breaking. In SQCD the
K\"ahler potential is uncalculable. However, considering a regime
where the K\"ahler potential is well approximated by the addition
of the leading order correction to the quadratic term, the
existence of uplifted vacua depends only on the sign in front of
this correction term.

We shall begin by assuming that the sign is the appropriate one,
and find the regime in parameters space leading to uplifted metastable states.
Then we shall use the uplifted vacua
as the local minima for direct gauge mediation of
SUSY breaking to the MSSM, and investigate some of their properties.

Deformed SQCD models,
of the type we shall study here, have a simple embedding on systems
of intersecting NS5-branes and D-branes in type IIA string theory \GiveonEW\
(for an introduction to such constructions, see e.g. \GiveonSR).
In the perturbative string regime, the corrections to the
K\"ahler potential for the spurion of SUSY breaking
(as well as higher derivative D-terms)
in the low-energy theory on the D4-branes are calculable \GiveonUR.
Similarly, one can calculate the  K\"ahler potential in the
T-dual theories consisting of D5-branes wrapped on
two-cycles in Calabi-Yau (CY) compactifications of the type IIB string theory \GMR.

The computations are done by considering the D-branes as small probes in the
background geometry of the NS5-branes (or the CY manifold)
thus providing, in particular, the K\"ahler potential.
Intriguingly, one finds the good sign!
(This is correlated with the simple fact that the gravitational
force felt by the D-brane probes in the background geometry is attractive
\refs{\GiveonFK,\GiveonEW}).
One may thus use such constructions in string theory as the hidden sector
for the direct mediation of SUSY breaking to the MSSM (the latter being realized as the
low-energy theory on some intersecting ``MSSM D-branes'' sector,
taken to be part of the ``flavor branes'' of the SUSY-breaking sector).

These brane constructions
include naturally matter in the adjoint of the low-energy gauge theory
(due to the rich structure of compact CY manifolds, or the necessity to
have many NS5-branes for the consistency of the four-dimensional type IIA theory).
The adjoint matter and its interactions enhance significantly the landscape of
uplifted vacua.
We shall leave the investigation of uplifted metastable vacua
in such string embeddings for future work.
Here we just present it as an interesting set of examples
where the K\"ahler potential in low-energy effective theories
is calculable and gives, in particular, the desired sign.

Motivated by the above discussion,
it is thus interesting to investigate the landscape of uplifted vacua
in deformed SQCD with adjoint matter fields,~\foot{Perturbatively
stable ISS-like vacua in SQCD with adjoint
matter were investigated e.g. in~\refs{\AmaritiVK,\CraigHF}.}
and its properties.
An initial study of this issue will be presented in this work.

Finally, given the rich landscape of generalized and deformed SQCD,
a natural question is: which of the local minima in the landscape is favored
during the cosmological evolution of the universe?
We shall address this question as well.
Concretely, we shall first assume that the reheating temperature is well above
the messenger scale, and that the universe cools down adiabatically.
This will allow us, as in
\refs{\AbelCR,\CraigKX,\FischlerXH,\AbelMY,\KatzGH},~\foot{In
\KatzGH\ it was argued e.g. that the uplifted vacua of \GiveonYU\
are disfavored cosmologically.}
to use the thermal potential as a tool to inspect the possible
phase transitions as the universe cools down, and to evaluate the likelihood
to end up in a phenomenologically interesting state in the landscape.

This note is organized as follows.
In section 2, we consider massive SQCD
in the free magnetic phase,
with a higher-dimension (``Planck'' suppressed) deformation
which leads to a tadpole term, $-aX$, in the scalar potential
for the spurion of SUSY breaking, $X$.
The latter pushes the system away
from the origin in pseudo-moduli space.
We assume that the sign in front of
the leading correction to the K\"ahler potential
is such that it can balance the tadpole ``repulsion,''
and find the conditions that the theory has long-lived {\it uplifted}
metastable vacua away from the tachyonic regime.

Moreover, we embed the MSSM in the flavor symmetry group of SQCD,
and require that the gaugino masses,
generated via direct gauge mediation, are at the weak scale.
We find that if the Yukawa coupling of the spurion to the magnetic quarks
is of order one, then the value of the tadpole is rather tuned.
However, if the Yukawa coupling is small,
then uplifted vacua satisfying all the requirements are generic.

In section 3, we inspect the thermal history of this model,
as the universe cools down from a high reheating temperature
towards the messenger scale.
We find that the cosmological evolution may favor uplifted vacua when the
tadpole term is relatively large.
However, we argue that one cannot find cosmologically favored vacua,
which satisfy all our conditions naturally
in pure (deformed) SQCD.
On the other hand,
as we show in section 4,
cosmologically favored
uplifted metastable states in viable models of gauge mediation
are generic in SQCD with adjoint matter.

Finally, we discuss some aspects of this work in section 5,
and present some technical details in an appendix.

\newsec{Uplifted vacua and gauge mediation in deformed SQCD}

\subsec{Gross understanding of the physics}

Consider deformed $SU(N_c)$ SQCD with $N_f$
flavors in the free magnetic phase \SeibergPQ\
(namely, $N_c<N_f<{3\over 2}N_c$;
for a review, see e.g. \IntriligatorCP).
In this subsection we are only interested in the gross understanding
of the physics, and thus we shall ignore
indices, traces, $N_{c,f}$ factors, etc.;
a more precise analysis will be done in the following subsection.

At low energies this theory is described by the Seiberg dual
``magnetic theory,'' which is IR free.
Concretely, in the IR we consider a theory with a superpotential
\eqn\www{W=h\left(X q\tilde q-\mu^2 X+\half\epsilon\mu X^2\right)~,}
and a non-trivial K\"ahler potential
\eqn\kkk{K=X\bar X-{\gamma\over\Lambda^2}(X\bar X)^2+\dots~,}
where $\Lambda$ is the QCD scale.
We choose the magnetic scale to be equal to the
electric one. The meson field $X$ is given in terms of the
electric quarks $Q,\tilde Q$ by (see \IntriligatorDD\ and the review
\IntriligatorCP)~\foot{If $\tilde\Lambda\neq\Lambda$, where $\tilde\Lambda$
is the magnetic scale, then $X={1\over\sqrt{\alpha}\Lambda}Q\tilde Q$,
where $\alpha$ is a real number of order one in the K\"ahler potential,
$K={1\over\alpha|\Lambda|^2}Tr M^{\dagger}M+\dots$, $M\equiv Q\tilde Q$,
and $h={\sqrt{\alpha}\Lambda\over\hat{\Lambda}}$, where
$\hat{\Lambda}$ is given in terms of $\Lambda$ and $\tilde\Lambda$ by the
scale matching $\Lambda^{3N_c-N_f}\tilde\Lambda^{3(N_f-N_c)-N_f}=\hat{\Lambda}^{N_f}$
(see e.g. \IntriligatorDD; here we ignore phases).}
\eqn\xqq{X={1\over h\Lambda}Q\tilde Q~,}
and the $N_f$ magnetic quarks $q,\tilde q$ are in the
fundamental and anti-fundamental of the magnetic gauge group
$SU(N_f-N_c)$. The constant $\gamma$ in \kkk\ is an uncalculable
real number. We {\it assume} here that $\gamma$ is {\it positive}.

We look for local minima in which $q=\tilde q=0$ and $X$ is proportional
to the identity matrix~\foot{In the notation of
\refs{\GiveonEF,\GiveonEW,\GiveonUR},
these are the vacua with $k=0$ and $n=N_f$. These are the most uplifted vacua.
Vacua with $k$ Higgsed magnetic quarks
have a larger hierarchy between the gaugino and sfermion masses \GiveonYU.
In particular, in the ISS-like vacua, $k=N_f-N_c$, the gaugino masses vanish to leading
order in SUSY breaking, even though R-symmetry is broken.},
and denote its common eigenvalue also by $X$.
To leading order, the potential is
\eqn\vvv{{1\over h^2}V=\left|\mu^2-\epsilon\mu X\right|^2\left(1+{4\gamma}{|X|^2\over\Lambda^2}+
{h^2\over 16\pi^2}\log\left(|X|^2\right)+\dots\right)+\dots~,}
where the dots stand for higher order corrections in ${X\over\Lambda}$
and in the loop expansion parameter ${h^2\over 16\pi^2}$,
and for non-perturbative corrections.
{}From now on we shall take $\epsilon,h,\mu$ and $\Lambda$ to be real and positive.

A few comments are in order:
\item{(i)}
In the UV, the quadratic term in \www\ is quartic in the electric
quarks, and hence suppressed by the high-energy scale $M_*$ (which
can be taken to be e.g. the Planck scale):
\eqn\xxqqqq{h\epsilon\mu X^2\simeq {1\over M_*}(Q\tilde Q)^2~.}
Consequently,
\eqn\eee{h\epsilon\simeq{(h\Lambda)^2\over \mu M_*}~.}
\item{(ii)}
The mass squared of the magnetic squarks at the minimum is (see appendix A)
\eqn\mmq{m^2_{q\pm}=h^2\left(X^2_{min}\pm\mu^2+O(\epsilon)\right)~,}
and thus some of the squarks are tachyonic if $X_{min}$
is not sufficiently large. We are thus looking for a local minimum at
\eqn\xmin{X_{min}>\mu~.}
\item{(iii)}
{}For $X\gg\mu$ it is justified to use the log approximation for the
one-loop Coleman-Weinberg potential
(see appendix A), as was done in eq. \vvv.
Although we only require $X_{min}$ to satisfy \xmin,
we shall still use the log approximation since we are only interested
in a gross understanding of the physics.

Since we take all the parameters to be real and
positive, we can treat $X$ as real. We also take~\foot{If, instead,
$\epsilon\mu>\mu^2/\Lambda$, such that the correction to $K$ is negligible,
the general argument in appendix A of \GiveonYU\
implies that the log term from one-loop prevents the existence of a local minimum
at $X$ which is bigger than the largest mass scale in $V$, $\epsilon\mu$.
So in this case there is no local minimum \xmin.}
\eqn\mule{\epsilon<\sqrt{4\gamma}\left(\mu\over\Lambda\right)~,}
so the potential is approximated by~\foot{Strictly speaking,
the potential is well approximated by this equation only when
$\epsilon\ll\sqrt{4\gamma}\left(\mu\over\Lambda\right)$,
but we are only interested in the gross understanding
of the physics, for which \mule\ is fine.}
\eqn\vsim{V\sim \mu^4-2\epsilon\mu^3 X+{4\gamma}\mu^4{X^2\over\Lambda^2}
+2\left(h\over 4\pi\right)^2\mu^4\log
X+\dots~.}
Thus, a local minimum exists if
\eqn\exist{X>{h\over 4\pi}{\Lambda\over\sqrt{4\gamma}}~,} and if
\eqn\roughly{\epsilon>\sqrt{16\gamma}{h\over 4\pi}{\mu\over\Lambda}~.}
Its location is at
\eqn\findmin{X_{min}\simeq{\epsilon\Lambda^2\over 8\gamma\mu}~.}
In order for the magnetic description to be valid at $X_{min}$ we
must have
\eqn\xminl{X_{min}\ll\Lambda~.}
To estimate the lifetime of the metastable vacuum,
we need to compute the Euclidean action
which gives the amplitude of tunneling to a lower minimum.
There are a few cases all of which are discussed next.

Consider first the bounce action for tunneling to a true minimum.
Approximating the barrier by a triangular potential \duncan, one finds that
\eqn\sss{S_{susy}\sim {X_{susy}^4\over V_{min}}\simeq{X_{susy}^4\over
h^2\mu^4}~.}
There are two types of cases now.
If $\mu\over\epsilon$ is relatively large,
then
the location of the supersymmetric minima is dictated by the
non-perturbative superpotential to be:
\eqn\xsusy{X_{susy}\simeq{\Lambda\over
h}\left({\mu\over\Lambda}\right)^{2(N_f-N_c)\over N_c}~,}
so metastability requires
\eqn\metareq{{1\over h^6}\left(\Lambda\over\mu\right)^{4(3N_c-2N_f)\over N_c}\gg 1~,}
which is easily satisfied in the free magnetic phase
(if $N_f/N_c$ is not too close to $3/2$).
On the other hand, if $\mu\over\epsilon$ is relatively small,
then
\eqn\xsusyb{X_{susy}={\mu\over\epsilon}~,}
and metastability requires
\eqn\metareqb{{1\over h^2\epsilon^4}\gg 1~,}
which is usually true, \eee.

Similarly, the bounce action for tunneling to a lower metastable minimum
is
\eqn\smet{S_{meta}\sim {X_{min}^4\over h^2\mu^4}~,}
since the distance between two local minima which differ by the
number of Higgsed quarks is roughly $X_{min}$, and $\Delta V\simeq h^2\mu^4$.
Hence, metastability requires that either $X_{min}$ is sufficiently bigger than
$\mu$ and/or $h^2\ll 1$:
\eqn\metareqc{{1\over h^2}\left({X_{min}\over\mu}\right)^4\gg 1~.}
Finally, we should also make sure that
\eqn\xminxsusy{X_{min}<X_{susy}~,}
such that, in particular, non-perturbative contributions
are negligible near the SUSY-breaking vacuum,
for the estimation \findmin\ to be valid.

Next we will embed the MSSM in
the $SU(N_f)$ flavor symmetry of the model above and discuss the
direct mediation of SUSY breaking~\foot{The hidden sector thus has an unbroken
$SU(N_f-N_c)$ gauge symmetry and consequently light particles;
these are very weakly coupled to the MSSM.
{}For $N_f-N_c\neq 1$, the dynamics of these light d.o.f.
should be taken into account, but below,
we will usually consider the case $N_f-N_c=1$.}
(for a review, see e.g. \GiudiceBP).
The spurion for SUSY breaking is $X$ and the messengers are (some of) the
$q,\tilde q$.~\foot{Taken to be e.g. in the $5,\bar 5$ of $SU(5)$ GUT.}
To leading order in SUSY breaking, the masses of the gauginos are
\eqn\gaugino{m_{\lambda_r}={\alpha_r\over 4\pi}\Lambda_G~,}
where
\eqn\lambdag{\Lambda_G\equiv{F_X\over X_{min}}\simeq{h\mu^2\over X_{min}}~.}
The mass squared of the sfermions is
\eqn\sfermion{m^2_{\tilde f}=2\sum_{r=1}^3 C_{\tilde f}^r\left({\alpha_r\over 4\pi}\right)^2\Lambda_S^2~,}
where the ratio of $\Lambda_S$ and $\Lambda_G$ is given in terms of the messenger's number
$N_{mess}$:~\foot{Definitions of $\Lambda_G$, $\Lambda_S$
and an effective number of messengers, as well as its notion for theories that respect
just the SM gauge symmetry $SU(3)_C\times SU(2)_W\times U(1)_Y$,
in a general framework, are given in~\CheungES,
but we will not need these here.}
\eqn\lgls{N_{mess}={\Lambda_G^2\over\Lambda_S^2}~.}
Note that although the mediation of SUSY breaking to the MSSM is direct,
technically we have here a Minimal Gauge Mediation.~\foot{This
is not true in vacua where $k$ of the quarks are Higgsed. In particular,
in the latter vacua, $N_{mess}$ \lgls\ will be smaller than the number of messengers.
Actually, it will even be smaller than the number
of messengers which contribute to the gaugino masses,
as explained in \GiveonYU.}

To obtain gaugino masses near $100$ GeV we should thus require
that
\eqn\require{\Lambda_G\simeq{8\gamma h\mu^3\over\epsilon\Lambda^2}\simeq
10^5\,\,{\rm GeV}~.}
Together with \eee\ this implies
\eqn\mulambda{{\mu\over\Lambda}\simeq\left({10^5\,{\rm GeV}\over 8\gamma M_*} \right)^{1\over 4}~.}
Taking, for example, $h,\gamma\simeq 1$,  which are their natural
values in SQCD, and $M_*\simeq M_{Planck}$,
we want to satisfy all the constraints
and requirements. The system is highly constrained, but has solutions in parameters space,
for instance:
${\mu\over\Lambda}\simeq 10^{-3.5}$, $\epsilon\simeq
{1\over 3}{\mu\over\Lambda}$, $\Lambda\simeq 10^{10.5}$ GeV,
$X_{min}\simeq{\Lambda\over 25}<X_{susy}\simeq\Lambda/10^{7(N_f-N_c)\over N_c}$ is satisfied
if $N_f<{6\over 5}N_c$. In this case, the bounce action \sss\ is
large, \metareq, as well as the bounce action \smet,
so longevity is secured.

We shall end this subsection with a few comments:
\item{(1)}
We have verified numerically, by using the exact tree-level and one-loop potentials,
that indeed minima of this type are easily obtained;
this is described in the next subsection.
\item{(2)}
A particularly interesting case is the case of a much larger tadpole term.
A sufficiently large $\epsilon$ will allow the cosmological evolution
to favor an uplifted vacuum, as we shall see later.
Consider for instance $\epsilon\simeq 1/10$.
Equation \mule\ implies that in this case $\mu/\Lambda$
cannot be much smaller than $1/10$ as well.
The approximations after eq. \mule\ show that
one can find in this case a local minimum.
We have verified numerically that uplifted minima
may exist in this range (see the next subsection).
However, now the condition \mulambda\ cannot be satisfied for a large $M_*$.
This implies that for a relatively large tadpole term,
uplifted vacua satisfying all our constraints
are impossible in pure SQCD.
However, by considering SQCD with an adjoint,
we will see that the landscape of possibilities is much richer,
in particular allowing viable uplifted vacua
which are favored cosmologically.
\item{(3)}
Finally, recall that
everything done above assumes that the constant $\gamma$ is positive.
In SQCD we do not know how to calculate $\gamma$.
However, by embedding the theory above on brane systems in string theory,
and going to the perturbative string regime,
one can calculate the K\"ahler potential (as well as higher derivative D-terms).
This was done in \GiveonUR,
where it was found that $\gamma$ is positive,
giving rise to the uplifted metastable vacua of the type analyzed above.

\subsec{More detailed analysis}

We shall now present a more precise analysis.
The superpotential is
\eqn\wwww{W=h\left(X_i^{\,j}q_a^i\tilde q_j^a-\mu^2 Tr(X)+{1\over 2}\epsilon\mu Tr(X^2)\right)~,}
where $i,j=1,\dots,N_f$ are flavor indices, $Tr$ is over flavor indices,
and $a=1,\dots,N$,
\eqn\nnnnn{N\equiv N_f-N_c~,}
is a color index of the magnetic gauge group.~\foot{We
may also add a double-trace deformation $(Tr(X))^2$ to \wwww\
which, in particular, gives different masses
to the adjoint and singlet of $SU(N_f)$.
This is sometimes a necessity, e.g. in order to
give sufficiently large masses to the adjoint fermions from
the $X$ superfield \EssigKZ\ and to push
the Landau-pole well above $\Lambda$ (as e.g. in \GiveonYU).
This term does not affect our gross understanding of the physics,
but it will be interesting to add it when inspecting the whole
features of our models.
}

The K\"ahler potential is
\eqn\kkkk{K=Tr(X^{\dagger}X)-{c_1\over\Lambda^2}Tr(X^{\dagger}X)^2
-{c_2\over\Lambda^2}(Tr(X^{\dagger}X))^2+\dots~,}
where the dots
stand for more terms compatible with the $U(N_f)$ symmetry and
higher order terms in $X/\Lambda$.~\foot{The terms omitted in \kkkk\
have negligible contributions. Similar corrections
to $K$ were considered for $N_f=N_c$ SQCD in \KatzGV.}
We look for solutions of the form
\eqn\xdiag{X_i^j=X\delta_i^j~,}
with
\eqn\xreal{X\in R_+~.}
Namely, we want to find local minima for the potential
\eqn\vavb{\eqalign{V=V_0+V_1+\dots=
&h^2N_f\left(1+4\gamma{X^2\over\Lambda^2}\right)\left(\mu^2-\epsilon\mu
X\right)^2\cr +{h^4N_fN\over 32\pi^2} &\Big[(X^2+\mu^2-\epsilon\mu
X)^2 \log\left(h^2(X^2+\mu^2-\epsilon\mu
X)\over\Lambda^2\right)\cr &+(X^2-\mu^2+\epsilon\mu X)^2
\log\left(h^2(X^2-\mu^2+\epsilon\mu X)\over\Lambda^2\right)\cr
&-2X^4\log\left(h^2X^2\over\Lambda^2\right)\Big] +h^2{\cal
O}(h^4,h^2(X/\Lambda)^2,(X/\Lambda)^4)~.}}
This potential is written on $q=\tilde q=0$, \xdiag,\xreal, $X<\mu/\epsilon$ and
\eqn\hme{h,\mu,\epsilon,\Lambda\in R_+~.}
The tree-level potential
$V_0=\left({\partial\over\partial{X_i^j}}{\partial\over\partial\bar
X_{\bar i}^{\bar j}}K\right)^{-1} {\partial W\over\partial
X_i^j}{\partial\bar W\over\partial\bar X_{\bar i}^{\bar j}}$
is obtained from \wwww\ and \kkkk, and the one-loop potential is
discussed in appendix A. The real parameter $\gamma$ is a certain
combination of the $c$'s in \kkkk,
\eqn\gammacc{\gamma=c_1+N_f c_2+\dots~,}
which we assume to be positive.

At large $X$, the one-loop potential can be approximated by:
\eqn\voneapp{ V_{1} \approx {h^4 N N_f \mu^4\over 8 \pi^2} \log X \, .}
If we add this to the leading terms (of order $X$ and $X^2$)
of the tree-level piece in \vavb, then
in order to have a minimum one gets:
\eqn\roba{ \epsilon >  {h  \sqrt{N \gamma}\over \pi } {\mu\over \Lambda}\, , }
and
\eqn\robb{ X_{min}\approx{\epsilon \Lambda^2\over 8 \gamma \mu}~.}
Another approximation can be used to estimate the upper bound for $\epsilon$.
When $\epsilon$ is sufficiently larger than the bound \roba,
the location of $X_{min}$ is large enough to ignore the
$\log X$ term in $V$.
So, now we ignore the one-loop term, but we keep
the whole tree-level potential in \vavb.
In order to have a minimum we get:
\eqn\robc{ \epsilon<\sqrt{{\gamma\over 2}}{\mu\over\Lambda}~,}
and
\eqn\robd{  X_{min}\approx{\mu\over 4 \epsilon}~.}
To summarize, in order to have a minimum we have to choose $\epsilon$ in the window
\eqn\robe{{h  \sqrt{N \gamma}\over\pi } {\mu\over\Lambda} < \epsilon
 < \sqrt{\gamma\over2}{\mu\over\Lambda} \, ,}
and the corresponding window for the location of the mimimum is:
\eqn\robf{{h\over 8 \pi } \sqrt{N\over\gamma}\Lambda<   X_{min}
< {1\over\sqrt{8 \gamma}} \Lambda  \, .}
A couple of comments are in order:
\item{(a)}
We see that for $h\sim 1$, which is its natural value in SQCD,
and for small $N$,~\foot{We
will usually restrict to the case $N=1$,
for which the dynamics of the unbroken $SU(N)$ gauge theory is trivial
(see also footnote 8). Yet, we shall keep the explicit $N$-dependence in the
equations below.}
the value of the tadpole $\epsilon$ must be rather tuned to have a local minimum.
Consequently,
the corresponding location of the minimum is highly restricted and, in particular,
its location is very far from the origin.
On the other hand, if $h\ll 1$, there is a large range of $\epsilon$ for which
a minimum exists, and thus uplifted vacua are generic.
Moreover, the corresponding location of the minimum is in a wide range and, in particular,
can be close to $X_{tachyon}\approx\mu$.
\item{(b)}
In order to trust the calculation we should  impose
\eqn\robg{{X_{min}^2\over\Lambda^2}\ll{1\over 4 \gamma} \, ,}
such that our approximation for $K$ is valid.
So we can just marginally trust the values of $\epsilon$
near the upper bound of the window \robe,\robf.

Taking into account the number of messengers, $N$, the scale $\Lambda_G$ in \gaugino,\lambdag\ is
\eqn\nlg{\Lambda_G\simeq{Nh\mu^2\over X_{min}}~.}
Thus, when $\epsilon$ is close to its smaller bound, from \robb\ we have
\eqn\erequire{
\Lambda_G\simeq{8N\gamma h\mu^3\over\epsilon\Lambda^2}\simeq 10^5\,\,{\rm GeV}~,}
which leads following \eee\ to
\eqn\exmulam{{\mu\over\Lambda}\simeq\left({10^5\,{\rm GeV}\over 8\gamma NM_*}\right)^{1\over 4}~.}
In this case, all the constraints can be satisfied, e.g. by numbers similar to those
presented in the previous subsection; we have also verified this numerically by using the
full potential \vavb.

On the other hand, when $\epsilon$ is closer to its upper bound, from \robd\ we have
\eqn\exmulamb{\Lambda_G\simeq 4N\epsilon h\mu\simeq 10^5\,\,{\rm GeV}~,}
which leads together with the condition \robc, and the relation \eee,
to the bound
\eqn\mbound{M_*<{\gamma\over 8N\epsilon^4}10^5\,\,{\rm GeV}~.}
Consequently, in pure SQCD, if the tadpole $\epsilon$
is large the high-energy scale $M_*$ is anomalously small.

This will be discussed further in the next sections,
where we will argue, in particular, that this problem is cured
by adding adjoint matter to SQCD.
However, we shall first show that a large tadpole term
is necessary for the universe
to end its early evolution in an uplifted vacuum,
once the reheating temperature is well above the messenger scale.
This is the topic of the next section, where we
shall inspect the thermal potential in our deformed SQCD.

\newsec{The Thermal Potential}

We now turn to study the thermal history of the model in the
previous section. We will assume that after inflation and
reheating the universe is in a thermal state with a temperature
well above the messenger scale, and that it cools down
adiabatically. This allows us to investigate the evolution of
the universe from the end of reheating till a (metastable) vacuum is
populated, by inspecting the thermal effective potential.
The purpose of this section is to find the conditions allowing
the cosmological evolution to favor an uplifted metastable state,
as it cools down after inflation and reheating.

We shall thus begin by inspecting the thermal potential.
The expression for the one-loop effective thermal potential is
obtained by adding to the zero-temperature potential \vavb\ the term
\DolanQD:
\eqn\vterm{ V_{th}=\sum \left( \pm {T^4\over 2 \pi^2} \int_0^\infty
dw \, w^2 \, \log \left( 1 \mp e^{-\sqrt{w^2+ {m^2(X,q)\over T^2}}}\right)
\right)~,}
where the upper sign is for bosons, the lower sign for fermions,
and the sum is over all the degrees of freedom.
The masses of the various particles depend on the values of both
the pseudo-moduli $X$ and the magnetic quarks $q,\tilde q$.

Now, if the temperature is lower than the mass of the particles in
the system, the contribution is essentially zero.
On the other hand,
in the regime where $T$ is much bigger than the mass of the particles,
we can expand in $m^2/T^2$:
\eqn\vthapp{ V_{th} \approx -{\pi^2\over 90} T^4 \left( N_{\rm bos}+{7\over 8} N_{\rm fer}\right)
+ {T^2\over 24} \left( \sum_{\rm bos} m^2(X,q) +\sum_{\rm fer} {m^2(X,q)\over 2 }\right)~.}
We shall now look at the thermal potential in two cases:
on the ``mesonic direction,''
\eqn\mesdir{q=\tilde q=0~,\qquad X_i^j=X\delta_i^j~,}
where the magnetic quarks are not Higgsed,
and the ``squarks direction,''
\eqn\squarkdir{q=\tilde{q}^\dagger= \left(
 \matrix{
 q_0 {I}_{N \times N}   \cr
    0}
\right)\, ,  \qquad X=0~,}
where the $SU(N)$ gauge symmetry is broken.

On the mesonic direction, using the spectrum presented in appendix A,
one finds that
\eqn\vthmes{ V_{th}(X,q=0)\approx 0 \qquad {\rm for}  \qquad X\gg{T\over h}~,}
while~\foot{Strictly speaking, this is true only for $N=1$.
{}For bigger $N$, we should also take into account the effects of the
$SU(N)$ gauge dynamics, e.g. the contribution of
the magnetic gauge bosons and D-terms (see \FischlerXH). These give corrections
which are negligible in the limit of a small magnetic gauge coupling.
We also ignore the MSSM interactions for similar reasons
(see \AbelMY\ and \KatzGH).}
\eqn\vthmasb{ V_{th}(X,q=0)\approx
N_f N \left( -{\pi^2\over 12} T^4 + {T^2\over 4}  h^2 X^2 + \ldots \right)
\qquad {\rm for}  \qquad  X\ll{T\over h}~.}
Recall that the full thermal effective potential, $V(T)$, is obtained by adding
the $T$-dependent piece $V_{th}$ to the zero-temperature
potential \vavb:
\eqn\vtvova{V(T)\equiv V_0+V_1+V_{th}+\dots~.}
{}For high temperatures, $V(T)$ has a minimum at
\eqn\xth{X_{th}\approx{4 \epsilon \mu^3\over NT^2}~.}
This thermal minimum is thus very close to the origin, at very high temperatures,
but it is pushed away from the origin as the universe cools down.

This effect is clearly due to the existence of the tadpole term -- the $\epsilon$ term
in \vsim. Evolving into a thermal minimum, which is located away from the origin
as the universe cools down from $T_{reheat}$ towards the symmetry breaking scale,
thus requires a large tadpole term.

We now turn to the thermal potential in the squarks direction \squarkdir.
Using the spectrum presented in appendix A, one finds:
\eqn\vthq{V(T;X=0,q_0)\approx
h^2 \left( N(q_0^2-\mu^2)^2 + (N_f-N) \mu^4 \right)+
T^2 h^2 \left( {1\over 8} N_f^2 \epsilon^2 \mu^2 +{1\over 2} N N_f q_0^2\right)~,}
where $q_0$ is defined in \squarkdir.
Hence, at a sufficiently low temperature, squarks become tachyonic
and a second order phase transition, towards an ISS-like vacuum,
occurs.~\foot{The new basin of attraction at $X\gg T$ and $q_0\sim\mu$,
taken into account in \KatzGH, is irrelevant here
since the messengers masses, required for this to exist, behave like $q_0$ and
thus emerge only after the phase transition to the ISS-like vacuum already occurred.}

The critical temperature for the second order phase transition is:~\foot{This
is slightly different from the results of \FischlerXH, who get
(in the limit of small magnetic gauge coupling):
$T_c^2={12 \mu^2\over 3  N_f  -N}$.}
\eqn\tcrit{T_c^2\approx {4 \mu^2\over N_f}~,}
and the value of the thermal minimum \xth\ at this point is:
\eqn\robt{X_{min}(T=T_c)\approx{\epsilon \mu N_f\over N}~.}
The location of the thermal minimum as the universe reaches the critical temperature
must be bigger than $\mu$, in order to avoid the phase transition to the ISS-like vacuum.
This requires the condition
\eqn\ebignf{\epsilon>{N\over N_f}~.}
Let us inspect this in more detail

Consider e.g. the case $N_f \approx 10$ and $N=1$, for which
the smallest tadpole allowed by the requirement \ebignf\ is $\epsilon \approx 1/10$.
In this case, the zero-temperature uplifted minimum is located at
$X_{min}\approx 2\mu$ (see eq. \robd), so the universe is likely
to slide towards this uplifted state as it continues to cool down.

There is however a serious drawback in this scenario.
{}From eq. \mbound\ we see that such a large tadpole
leads, in particular,
to an anomalously small high-energy scale, $M_*<10^8$ GeV,
and one cannot satisfy all our constraints~\foot{If $h$
is of order one, then the gauge mediation vacuum is
not metastable, \metareqc, while for small $h$,
eq. \exmulamb\ sets the symmetry-breaking
scale $\mu$ above the cutoff scale $M_*$.}.
We shall next cure this problem by considering
uplifted vacua in (deformed) SQCD with adjoint matter.

\newsec{Uplifted vacua in SQCD with an adjoint}

We shall now show
that since the landscape of uplifted vacua in SQCD with an adjoint is richer,
in particular, it allows the existence of viable uplifted metastable states in
the presence of a large tadpole term (large $\epsilon$),
that are favored by the cosmological evolution
even when the reheating temperature is well above the messenger scale.
In this work, we shall focus on a rather limited regime
in the landscape of possibilities. Hence, one should regard
our investigation in this section
only as the initial step towards a more thorough
study of uplifted metastable vacua in SQCD with an adjoint, and the
mediation of SUSY breaking to the MSSM.

Before adding the adjoint though, let us recall the problems with
having cosmologically favored uplifted vacua in pure (deformed) SQCD.
The electric mesons $M\equiv Q\tilde Q$ are related to the canonically
normalized IR fields $X$ via $M=\sqrt\alpha \Lambda X$,
where $\alpha$ is a positive real parameter of order one,
appearing in $K={1\over\alpha |\Lambda|^2}M^\dagger M+\dots$.
Now, when the magnetic scale is chosen to be equal to the electric scale,
then $h=\sqrt\alpha$, thus $h$ is also of order one,
and since $h\epsilon={\alpha\Lambda^2\over\mu M_*}$,
this cannot be satisfied unless the high-energy scale $M_*$ is very
small, in which case one cannot satisfy all our constraints.
And even if it turns out that $\sqrt\alpha=h\ll 1$, the relation
$h\epsilon={h^2\Lambda^2\over\mu M_*}$ is still problematic
(since the high-energy scale $M_*$ is way too small).

To be able to resolve these problems, we need a model
where the value of the Yukawa coupling $h$ is separated from
the value of the parameter $\alpha$.
For instance, if $h\ll 1$ is possible even
when $\alpha$ is order one,
we are able to satisfy all the constraints
in sections 2 and 3.

This is the situation in SQCD with an adjoint,
as we shall now see.
First we recall some results of Kutasov, Schwimmer and Seiberg
\refs{\KutasovVE,\KutasovNP,\KutasovSS}
(see also \CraigHF).
Consider an $N=1$ $SU(N_c)$ SYM with $N_f$ flavors $(Q,\tilde Q)$, an adjoint $\Phi$
and a superpotential
\eqn\wel{W_{el}=Tr\sum_{n=2}^{k+1}{g_n\over n}\Phi^n~.}
The dual magnetic theory is an $SU(N)$ SYM, where
\eqn\nnkk{N=kN_f-N_c~,}
with certain matter and a superpotential as follows.
{}For simplicity we shall consider, as before, the case
\eqn\none{N=1~,}
for which there is no magnetic gauge group.
In this case, the IR theory has $N_f$ flavors $(q,\tilde q)$
and $k$ meson fields $X_j$, which are given in terms of the UV variables by
\eqn\xxxj{X_j={1\over\sqrt\alpha_j\Lambda^{j+1}}Q\Phi^j\tilde Q~,\qquad j=0,1,\dots,k-1~,}
where $\Lambda$ is the dynamically generated scale
of the $SU(N_c)$ gauge theory and $\alpha_j$ are order one positive
numbers appearing in the K\"ahler potential
$K=\sum_{j=0}^{k-1} {1\over\alpha_j|\Lambda|^2}Tr(M_j^{\dagger} M_j)+\dots$,
$M_j\equiv Q\Phi^j\tilde Q$,
and there is a superpotential
\eqn\wmag{W_{mag}=\sum_{j=0}^{k-1}h_jX_jq\tilde q~.}
The IR fields $X_j, q, \tilde q$ are canonically normalized.

As before, we choose the magnetic scale to be equal to the electric
one.~\foot{The scale matching in SQCD with an adjoint is
$\Lambda^{2N_c-N_f}\tilde\Lambda^{2N-N_f}=(\hat\Lambda/g_{k+1})^{2N_f}$
(see \KutasovSS\ for the definitions of the various scales).}
The Yukawa couplings are functions of the dimensionless parameters of the theory
\eqn\hofpar{h_j=h_j(\alpha_i,g_n\Lambda^{n-3})~,}
whose precise values can be found in \KutasovSS.
The $h_j$ are thus free parameters, which can take in particular very small values,
even though the $\alpha_i$ are order one parameters.
This aspect is an important difference with respect to the situation in pure SQCD.

To break supersymmetry in uplifted vacua we deform the theory to~\foot{This is the minimal
deformation required for our purpose; we could have added more terms at this order,
but their effect will be negligible in what we shall do next.}
\eqn\wdef{W=\sum_{j=0}^{k-1}h_j\left(X_jq\tilde q-\mu_j^2X_j+{1\over 2}\epsilon_j\mu_jX_j^2\right)~,}
where various traces are understood and the $\epsilon_j$ terms are high-dimension terms
in the UV fields, and thus
\eqn\hheeaa{h_j\epsilon_j\simeq{\alpha_j\Lambda^{2j+2}\over\mu_jM_*^{2j+1}}~,}
with $M_*$ being a high-energy scale.
Unlike pure SQCD, now relatively large tadpole terms are easily
obtained, even for large $M_*$,
and consequently, cosmologically favored uplifted vacua
satisfying all the other constraints are generic
even when the reheating temperature is well above the messenger scale.

We shall now focus on a concrete example, the $k=2$ case.
In this case,
\eqn\weltwo{W_{el}=Tr\left({g\over 3}\Phi^3-{m_\Phi\over
2}\Phi^2\right)~,}
namely, $g\equiv g_3$ is a dimensionless coupling
and $m_\Phi\equiv -g_2$ is the mass of the adjoint. {}For
$N=\Lambda/\tilde\Lambda=\alpha_j=1$,
one finds (following \refs{\KutasovSS,\CraigHF}) in this case
\eqn\hoha{h_1={1\over g}~, \qquad h_0={N_c-1\over 2g^2}
{m_\Phi\over\Lambda}~.}
So, small $h_j$, and thus large
$\epsilon_j$, are possible,
and cosmologically favored uplifted vacua,
which satisfy all the constraints, can thus exist.

{}For instance, if $g$ is large,
then $h_1$ is small,
and since $h_0\ll h_1$, the $X_0$ sector is almost decoupled from
SUSY breaking. Consequently, the analysis of sections 2 and 3 applies.
In particular, if $\epsilon_1$ is sufficiently large,
and since $\alpha_i$ are order one,
the uplifted vacua are viable, cosmologically favored
gauge mediation minima.
Similarly, if SUSY breaking is almost entirely
in the $X_0$ sector, and if $h_0$ is  small
(e.g. if $m_\Phi\ll\Lambda$), then
for sufficiently large tadpole terms,
cosmologically favored uplifted vacua may exist.
{}From these examples,
it seems possible that this is generic in SQCD with an adjoint.

Finally, a comment is in order.
{}For small Yukawa couplings, $h_i\sim g_{SM}$,
one should also take into account
the effect of the MSSM interactions (see \AbelMY\ and \KatzGH).
It appears though that this effect will help in favoring the uplifted
metastable vacua, for the following reason.
The supersymmetric vacuum has more light particles than the non-supersymmetric ones
(even at finite temperature) because the degeneracy between the SM particles
and their superpartners is bigger, so this effect reduces the relative number
of light particles near the origin, and thus decreases the probability
for the second order phase transition to the ISS-like vacuum to occur
(e.g., it decreases the value of the critical temperature \tcrit).
This effect is bigger the larger the SM couplings are relative to our
Yukawa couplings, $h$. So in our case, of a relatively large tadpole $\epsilon$
and thus small $h$, it seems that this effect
increases the probability that the cosmological dynamics will favor the
uplifted metastable vacua.

\newsec{Discussion -- uplifted metastable vacua in the cooling universe}

The main result of this work can be summarized as follows.
Assuming that the sign in front of the leading correction to the
K\"ahler potential for the spurion of SUSY breaking $X$
is appropriate, we argued that
generalized SQCD -- concretely, deformed SQCD with an adjoint --
has generic, cosmologically favored, long-lived metastable vacua,
which can be used in viable models of direct gauge mediation.

Explicitly, for theories in the free magnetic phase,
when the Yukawa coupling of $X$ to the magnetic quarks is small,
a relatively large tadpole
term for $X$ is naturally generated (from ``Planck'' suppressed
high-dimension operators), giving rise to cosmologically favorable
uplifted metastable vacua.
By embedding the MSSM in the flavor group of this generalized SQCD,
and setting the messenger scale appropriately,
one can obtain phenomenologically interesting models.

After inflation and reheating,
as the universe cools down adiabatically towards the messenger scale,
the tadpole term pushes the thermal minimum away from the origin in $X$.
And when the tadpole term is sufficiently large,
the universe is likely to populate an uplifted metastable state,
of the type desired phenomenologically, before the second order phase
transition towards the undesired metastable ISS-like vacuum occurs.

On the other hand, when the tadpole term $\epsilon$ is small,
our study gives a bound on the reheating temperature, as in \KatzGH.
Namely, in this case $T_{reheat}$
must be smaller than the symmetry breaking scale,
such that the universe may be trapped in a phenomenologically viable
uplifted vacuum already during inflation.
This is a rather severe constraint for cosmology if we require a
low-scale gauge mediation of SUSY breaking.

As discussed in the introduction, the (generalized) deformed SQCD models
of the type considered here have simple embedding in string theory, e.g.
on systems of D4-branes stretched between NS5-branes,
and the effective theory for the spurion $X$  is remarkably
similar to the one assumed here.
Vacuum selection in such systems was analyzed recently
in \KutasovKB, by considering the dynamics of the D4-brane probes
in the background of {\it non-extremal} NS5-branes.
The analog of the cooling universe in this picture is decreasing
the non-extremal scale.

The results in \KutasovKB\ are compatible with  the results in this work.
Namely, when the tadpole term is sufficiently large, the dynamics select
uplifted metastable states.
We should emphasize that while the models studied here and in \KutasovKB\
are connected continuously in string theory \refs{\GiveonFK,\GiveonEW},
this is not sufficient for neither the early universe dynamics,
nor the zero temperature one to be identical.
Nevertheless, we do find striking similarities.

Actually, the theories investigated in \KutasovKB\ correspond only to the
embedding of (deformed) pure SQCD in the type IIA brane construction.
It will be interesting to repeat the analysis of \KutasovKB\ to the
brane embedding of SQCD with an adjoint.

Moreover, it should be interesting to investigate the
dynamical vacuum selection in models with generic $k$ and $N$
in section 4, and to consider more general theories,
e.g. including couplings of the adjoint $\Phi$ to the quarks $Q,\tilde Q$,
as well as adding more matter to the theory, both in gauge theory
and in its string embeddings.

Finally, it will also be interesting to use our uplifted vacua
in deformed SQCD with an adjoint, to reconsider aspects of flavor physics
along the lines of the recent work \CraigHF.
The authors of \CraigHF\ embedded the first and second SM generations
inside the composite fields $Q\Phi\tilde Q$ and $Q\tilde Q$ of section 4,
respectively, to provide an explanation
to the flavor hierarchy within a ``single-sector'' SUSY-breaking
model~\refs{\ArkaniHamedFQ,\LutyVR}.
However, they studied the theory in an ISS-like vacuum,
which suffers from split supersymmetry.
On the other hand, investigating such models in our uplifted metastable vacua
may lead to interesting flavor physics, in models that also provide an
explanation to the hierarchy between the weak scale and the Planck scale.

\bigskip
\noindent{\bf Acknowledgements:} We thank Andrey Katz,
Zohar Komargodski, David Kutasov and Tomer Volansky for discussions.
This work was supported in part
by the BSF -- American-Israel Bi-National Science Foundation,
by a center of excellence supported by the Israel Science Foundation
(grant number 1468/06), DIP grant H.52, and the Einstein Center at the Hebrew University.

\bigskip
\appendix{A}{Potentials and spectra}

The superpotential of the Seiberg dual of deformed
$SU(N_c)$ SQCD with $N_f$ flavors is:
\eqn\aaaa{W=h q^i X_i^j\tilde{q}_j + h \Tr \left( -\mu^2 X+{1\over 2} \epsilon \mu X^2 \right)~,}
where $i,j=1,\dots,N_f$, the meson superfields $X_i^j$ are singlets,
and the magnetic quarks $q^i$ ($\tilde q_j$) are in the (anti-)fundamental
of the magnetic gauge group
\eqn\aaaaa{SU(N)~,\qquad N\equiv N_f-N_c~.}
The tree-level potential, in the case of a canonical K\"ahler potential, is:
\eqn\aaab{V = |h|^2 \left( |q^i \tilde{q}_j-\mu^2 \delta^i_j +\epsilon \mu X^i_j |^2
+|X q|^2 + |\tilde{q} X|^2   \right)~.}
Next we present the tree-level spectrum and the one-loop effective potential \EssigKZ\
(see also \GiveonBV).

\subsec{Mesonic direction: $q=\tilde q=0$, $X_i^j=X\delta_i^j$}

The tree-level spectrum in this direction includes:~\foot{{}From now on we take
$\epsilon,h,\mu$ to be positive real numbers.}
$$ 4 N_f N \, {\rm fermions,} \qquad m^2=h^2 X^2 \, ,$$
$$ 2 N N_f \, {\rm scalars,} \qquad m^2 =h^2(X^2 +\mu^2-X \epsilon \mu ) \, , $$
$$ 2 N N_f \, {\rm scalars,} \qquad m^2 = h^2(X^2 -\mu^2+X \epsilon \mu ) \, , $$
The mass squared of the last group of scalars is negative for $X<\mu$.
There are also
$$ 2 N_f^2 \, {\rm scalars,} \qquad m^2=h^2 \epsilon^2 \mu^2 \, ,$$
$$ 2 N_f^2 \, {\rm fermions,} \qquad m^2=h^2 \epsilon^2 \mu^2 \, ,$$
whose total contribution to the one-loop potential is zero.
Their contribution to the thermal potential is also not important,
because it is independent of $X$.

The one-loop contribution reads:
\eqn\aaac{\eqalign{V_1 ={N_f N\over 32  \pi^2}
\left(  h^4(X^2 +\mu^2-X \epsilon \mu ) ^2 \log \left(
{h^2(X^2 +\mu^2-X \epsilon \mu )\over M_c^2}\right) +
\right. \cr
\left.+ h^4(X^2 -\mu^2+X \epsilon \mu ) ^2 \log \left( {h^2(X^2 -\mu^2+X \epsilon \mu)\over M_c^2}\right)
- 2   h^4 \, X^4   \log \left( {h^2 \, X^2\over M_c^2}\right)
\right) \, ,}}
where $M_c$ is a cutoff scale.
Changing the cutoff scale, $M_c\to\tilde M_c$, amounts to
a renormalization  of the overall constant of the tree-level potential,
\eqn\aaad{{h^4 N_f N \over 16 \pi^2} (\mu^2-\epsilon\mu X)^2
\log\left(\tilde M_c^2\over M_c^2\right) \, ,}
which is negligible.
In this work we took $M_c=\Lambda$.

\subsec{Squark direction: $X=0$}

We consider here the maximally Higgsed case (when $SU(N)$ is completely broken)
-- the ISS-like vacuum.
Using flavor symmetry, the VEV of the squarks can be chosen to be
\eqn\aaae{ q=\tilde{q}^\dagger= \left(
 \matrix{
 q_0 {I}_{N \times N}   \cr
    0}
\right)\, ,  \qquad X=0~.}
The potential along this direction is:
\eqn\aaaf{ V= h^2 \left( N(q_0^2-\mu^2)^2 + (N_f-N) \mu^4 \right)~.}
Next, let us present the classical spectrum.

There are the following scalars:
$$ 2 (N_f-N)^2 \qquad m^2 =h^2 \, \epsilon^2 \mu^2  \, , $$
$$ N^2 \qquad m^2 =\pm h^2 \, (\mu^2-q_0^2) \, , $$
$$ 2 N(N_f-N) \qquad m^2 = {h^2\over 2}
 \left(2
   q_0^2+\epsilon ^2 \mu ^2+\mu ^2  \pm \sqrt{4 q_0^2 \epsilon ^2 \mu ^2+\epsilon ^4 \mu ^4-2 \epsilon ^2 \mu ^4+\mu ^4}\right) \, , $$
$$ 2 N (N_f-N) \qquad m^2 = {h^2\over 2} \left(2
   q_0^2+\epsilon ^2 \mu ^2-\mu ^2
\pm\sqrt{4 q_0^2 \epsilon ^2 \mu ^2+\epsilon ^4 \mu ^4+2 \epsilon ^2 \mu ^4+\mu ^4}\right)
\, , $$
$$ N^2  \qquad m^2 = {h^2\over 2} \left(3   q_0^2 +\epsilon ^2 \mu ^2+\mu ^2
\pm \sqrt{q_0^4+10 q_0^2 \epsilon ^2 \mu ^2-2 q_0^2 \mu ^2+\epsilon ^4 \mu^4-2 \epsilon ^2 \mu ^4+\mu ^4}\right)
\, , $$
$$ N^2 \qquad m^2 = {h^2\over 2}
 \left(5 q_0^2 +\epsilon ^2 \mu ^2-\mu ^2
 \pm\sqrt{q_0^4+6 q_0^2 \epsilon ^2 \mu ^2-2 q_0^2 \mu ^2+\epsilon ^4 \mu ^4+2
   \epsilon ^2 \mu ^4+\mu ^4}\right)
\, . $$

Finally, there are the following fermionic degrees of freedom:
$$ 2 N^2  \qquad m^2 =0  \, , $$
$$ 2 (N_f-N)^2 \qquad m^2 =h^2 \, \epsilon^2 \mu^2  \, , $$
$$ 4 N (N_f-N) \qquad m^2=
{h^2\over 2} \left( 2 q_0^2+\epsilon ^2 \mu ^2
\pm\epsilon  \mu  \sqrt{4 q_0^2+\epsilon ^2 \mu ^2}\right)
 \, , $$
$$ 2 N^2  \qquad m^2=
 {h^2\over 2} \left(  4 q_0^2+\epsilon ^2 \mu ^2
 \pm \epsilon  \mu  \sqrt{8 q_0^2+\epsilon ^2 \mu ^2}
\right)  \,  . $$

\listrefs
\end